\documentstyle[psfig]{article}

\def\ltap{\raisebox{-.4ex}{\rlap{$\sim$}} \raisebox{.4ex}{$<$}}

\newcommand{\Rsl}{{\not  \!{R}}}

\def\bra {\langle}
\def\ket {\rangle}

\def\b {\beta}

\def\r {\rightarrow}

\def\dphi {\Delta \phi}
\def\dalfa {\Delta \alpha}
\def\ljpsi {\lambda_{J/\psi}}
\def\brpsik {B_d\rightarrow J/\psi K_S}
\def\bbrpsik {\bar{B}_d\rightarrow J/\psi K_S}
\def\tpsik {B_d (t)\rightarrow J/\psi K_S}
\def\tbarpsik {\bar{B}_d (t)\rightarrow J/\psi K_S}
\def\brpsikpl {B^+\rightarrow J/\psi K^+}
\def\brpsikmi {B^-\rightarrow J/\psi K^-}
\def\brpsikplmi {B^\pm\rightarrow J/\psi K^\pm}
\def\brphik {B_d\rightarrow \phi K_S}
\def\bbrphik {\bar{B}_d\rightarrow \phi K_S}

\def\brphikmi {B^-\rightarrow \phi K^-}
\def\brphikplmi {B^\pm\rightarrow \phi K^\pm}
\def\apsi {A_{J/\psi}}
\def\aphi {A_{\phi}}
\def\acp {a_{\rm CP}}
\def\acppl {a_{\rm CP}^+}
\def\acpdir {a_{\rm CP}^{\rm d}}
\def\acpmix {a_{\rm CP}^{\rm m}}

\def\bar {\overline}
\def\mslsq {m^2_{\tilde{e}_{iL}}}
\def\msnsq {m^2_{\tilde{\nu}_{iL}}}

\def\be {\begin{equation}}
\def\ee {\end{equation}}
\def\bea {\begin{eqnarray}}
\def\eea {\end{eqnarray}}
\def\n {\nonumber}
\def\bc {\begin{center}}
\def\ec {\end{center}}

\title{
\begin{flushright}
\small 
SINP/TNP/01-10\\
{\tt hep-ph/0105057} 
\end{flushright}
{\bf Can {\boldmath $R$}-parity violation 
lower {\boldmath $\sin 2\beta$?}}
}

\author{
{\sf Gautam Bhattacharyya $^{a,}$}%
\thanks{Electronic address: gb@theory.saha.ernet.in}~, 
{\sf Amitava Datta $^{b,}$}%
\thanks{Electronic address: adatta@juphys.ernet.in. On leave of absence
from Jadavpur University, Kolkata 700032, India.} ~~and  
{\sf Anirban Kundu $^{c,}$}%
\thanks{Electronic address: akundu@juphys.ernet.in} 
\\ [2.5mm]
$^a$ {\small Saha Institute of Nuclear Physics, 1/AF Bidhan
Nagar, Kolkata 700064, India}\\ 
$^b$ {\small Department of Physics,
Visva-Bharati, Santiniketan  731235, India} \\ 
$^c$ {\small Department of Physics, Jadavpur University, 
Kolkata 700032, India}
}

\date{}

\begin{document}

\maketitle

\begin{abstract}
Recent time-dependent CP asymmetry measurements in the $B_d\rightarrow
J/\psi K_S$ channel by the BaBar and Belle Collaborations yield
somewhat lower values of $\sin 2\beta$ compared to the one obtained
from the standard model fit. If the inconsistency between these
numbers persists with more statistics, this will signal new physics
contaminating the $B_d\rightarrow J/\psi K_S$ channel, thus disturbing
the extraction of $\beta$. We show that the $R$-parity-violating
interactions in supersymmetric theories can provide extra new phases
which play a role in significantly reducing the above CP asymmetry,
thus explaining why BaBar and Belle report lower values of $\sin
2\beta$. The same couplings also affect the $B_d\rightarrow \phi K_S$
decay rate and asymmetry, explain the $B\rightarrow\eta' K$ anomaly,
and predict nonzero CP asymmetry in dominant $B_s$ decays. The
scenario will be tested in the ongoing and upcoming $B$ factories.

\vskip 5pt \noindent
\texttt{PACS number(s)}: 11.30.Er, 13.25.Hw, 12.60.Jv, 11.30.Fs \\ 
\noindent
\texttt{Keywords:} CP violation, $B$ decays, supersymmetry, $R$-parity
violation
\end{abstract}

\vskip 20pt  

\setcounter{footnote}{0}
\renewcommand{\thefootnote}{\arabic{footnote}}
Long after its discovery in the $K$ system, evidence of CP violation
is now being substantiated also in the $B$ system, in particular, via
the CP asymmetry measurement in the `gold-plated' $\brpsik$ channel
\cite{sanda}. The CP asymmetry in the above channel is proportional to
$\sin 2\beta$ in the standard model (SM), where $\beta = {\rm Arg}~
(V^*_{td})$ is an angle of the unitarity triangle of the
Cabibbo-Kobayashi-Maskawa (CKM) matrix. Any non-zero determination of
$\beta$ would be a signal of CP violation. The BaBar and Belle
Collaborations, operating at the asymmetric $B$ factories at SLAC and
KEK respectively, have recently reported
\bea
\label{bfac} 
\sin 2\beta =
\left\{
 \begin{array}{ll} 
0.34 \pm 0.20 \pm 0.05 ~~({\rm BaBar ~\cite{babar}}), \\
0.58 {}^{+0.32}_{-0.34} {}^{+0.09}_{-0.10} ~~({\rm Belle ~\cite{belle}}). 
\end{array}
 \right. 
\eea 
When these are combined with the previous measurements,
namely, $\sin 2\beta = 0.84 \pm 1.05$ (ALEPH \cite{aleph}) and $\sin
2\beta = 0.79 \pm 0.44$ (CDF \cite{cdf}), the global average reads 
\bea 
\label{global}
\sin 2\beta = 0.48 \pm 0.16.  
\eea
On the other hand, using the experimental constraints from the
measurement of $|\epsilon|$, $|V_{ub}/V_{cb}|$, $\Delta m_d$, and
from the limit of $\Delta m_s$, the fitted value of $\sin 2\beta$,
{\em strictly} within the framework of the SM, has been obtained as
\be 
\label{smfit}
\sin 2\beta = 
\left\{
 \begin{array}{ll} 
0.75 \pm 0.06 ~\cite{parodi},\\
0.73 \pm 0.20 ~\cite{alilondon}.
\end{array}
 \right.  \ee 
The two numbers in Eqs.~(\ref{global}) and (\ref{smfit}) are at
present consistent within errors, though their central values, as it
should be noted, are fairly seperated. This separation may turn out to
be an ideal new physics hunting ground. With an expected reduction of
uncertainties of the parameters that go into the SM fit, and with an
improved determination of the CP asymmetry in $\brpsik$ as the
statistics accumulates\footnote{The BaBar goal is to bring down the
accuracy of $\sin 2\b$ measurement to $\pm 0.06$ with 30 fb$^{-1}$
data \cite{babar-hb}.}, the inconsistency between Eqs.~(\ref{global})
and (\ref{smfit}) may persist or may become even more prominant. In
that case, a possible intervention of new physics with a new phase
affecting the CP asymmetry in the $\brpsik$ channel cannot be ignored.

\vskip 5pt

In the SM, the asymmetry in the $\brpsik$ channel is almost entirely
mixing-induced. The time-dependent CP asymmetry is proportional to
$\sin 2\beta$, where $\beta$ appears in the phase of $B_d$-$\bar{B}_d$
mixing.  The decay amplitude of $\brpsik$ does not carry any weak
phase at leading order in the Wolfenstein parametrization.  New
physics might change the scenario in two ways. It can add a new weak
phase in $B_d$-$\bar{B}_d$ mixing and/or it can generate a new diagram
for $b \r c\bar{c}s$ decay amplitude that carries a new weak
phase. Since the decay is Cabibbo-favoured, one usually tends to
overlook the latter possibility. In this paper we examine the
situation in which the decay is indeed affected by new physics, but
the mixing amplitude is not. The point to note is that this scenario
also induces a new physics amplitude in $\brpsikpl$. Now two things
can happen. First, direct CP asymmetry may be induced\footnote{An
observation of direct CP violation even at a few percent level will
constitute a definite signal of new physics.} in both $\brpsik$ and
$\brpsikpl$, which is non-existent in the SM. Indeed, for this to
happen there must also exist a strong phase difference between the SM
and new physics diagrams. Second, the mixing-induced CP asymmetry now
depends not only on $\beta$, it involves a new weak phase as well. As
a result, equating the CP asymmetry to $\sin 2\beta$ would be
misleading.  A combination of the `true' $\beta$, the angle of the
unitarity triangle, and other new parameters should now be related to
the experimental CP asymmetry in the $\brpsik$ channel. This way it
may be possible to explain why Eqs.~(\ref{global}) and (\ref{smfit})
may disagree.

\vskip 5pt

Since the values of $\sin 2\b$ extracted from the CP asymmetry
measurements in the asymmetric $B$ factories have a tendency to be
somewhat lower than the SM fit value, one is prompted to look for
models where such tendency is favoured. In models of minimal flavour
violation, where there are no new operators beyond those in the SM and
no new weak phase beyond the one in the CKM matrix, a conservative
scanning of all relevant input parameters in a standard CKM-like
analysis yields an absolute lower bound on $\b$. The SM, several
versions of the minimal supersymmetric standard model (MSSM), and the
two Higgs doublet models are examples of this class. The most
conservative lower bound, as noted by the authors of
Ref.~\cite{buras-buras,buras-fl} is\footnote{The bounds are
conservative in the sense that they have been obtained by 
independently scanning all parameters under consideration.}
\be 
\label{l-bound}
\sin (2\b)_{\rm min} = 
\left\{
 \begin{array}{ll} 
0.42 ~({\rm Present}),\\
0.48 ~({\rm Future}).
\end{array}
 \right.  \ee  
In particular, it has been shown in \cite{buras-buras} that
in the MSSM and in the minimal Supergravity (mSUGRA) models, the
minimum $\sin 2\b$ are 
\be 
\label{mssm-bound}
\sin (2\b)_{\rm min} = 
\left\{
 \begin{array}{ll} 
0.40~(0.49) ~({\rm MSSM}),\\
0.53~(0.62) ~({\rm mSUGRA}),
\end{array}
 \right.  \ee  
where the numbers within brackets correspond to future measurements of
input parameters. In the context of supersymmetry, if CP violation in
the $K$ system is purely supersymmetric, i.e. $\epsilon$ and
$\epsilon'/\epsilon$ are completely explained by new phases of
supersymmetric origin, then the CKM phase will be constrained from the
charmless semileptonic $B$ decays and $B_d$-$\bar{B}_d$
mixing. Only in that case, as the authors of Ref.~\cite{masiero} have
argued, the CKM phase could be quite small leading to a very low
$\acp$. The implications of a low $\acp$ in the context of a generic
new physics scenario have been discussed in Refs.~\cite{nir1,nir2}.

\vskip 5pt

The thrust of this paper is to examine the r\^{o}le of supersymmetry
with broken $R$-parity \cite{rpar} in the context outlined above. This
brand of supersymmetry does not fall into the minimal flavour
violating class, as it introduces new tree-level flavour changing
operators. The essential points are outlined below. Recall that in the
MSSM gauge invariance ensures neither the conservation of lepton
number ($L$) nor that of baryon number ($B$). Using these quantum
numbers $R$-parity is defined as $R = (-1)^{(3B+L+2S)}$, where $S$ is
the spin of the particle.  $R$ is +1 for all SM particles and $-1$ for
their superpartners. In a general supersymmetric model one should in
principle allow $R$-parity-violating ($\Rsl$) interactions.  Tight
constraints on the strength of these interactions exist in the
literature \cite{review}.  Even though any concrete evidence for the
existence of $\Rsl$ terms is still lacking, the observation of
neutrino masses and mixings in solar and atmospheric neutrino data
suggests that it would be premature to abandon the $L$-violating
interactions \cite{rp-nu}. Indeed, to avoid rapid proton decay one
cannot simultaneously switch on both $L$- and $B$-violating
interactions and for this reason we impose $B$ conservation by
hand. The $L$-violating superpotential that we consider here is
$\lambda'_{ijk} L_i Q_j D^c_K$, where $L_i$ and $Q_i$ are lepton and
quark doublet superfields, and $D_i^c$ is down quark singlet
superfields. The essential point is that the $\lambda'$-interactions
can contribute to non-leptonic $B$ decays at the tree level via
slepton/sneutrino mediated graphs. In this paper we focus on $\Rsl$
effects arising at the $\brpsik$ decay amplitude level rather than
through $B_d$-$\bar{B}_d$ mixing\footnote{Indeed, we respect the
constraints on $\lambda'$ couplings from $\Delta m_d$.}. 

\vskip 5pt

A detailed analysis of $R$-parity violation on $\brpsik$ and $\brphik$
decay modes have earlier been carried out in Ref.~\cite{guetta}.  We
have extended the formalism of \cite{guetta}, which was based on
mixing induced CP violation only, by incorporating the effects of the
strong phase difference between the interfering amplitudes (an
essential ingredient of direct CP violation). The latter constitutes
the main source of CP violation in charged $B$ decays and is present
even in neutral $B$ decays. The current availability of CLEO data on
charged $B$ decays together with the isospin symmetry enable us to
extract quantitative results on the strong phase difference within
this generalised framework. We also find that our essential conclusion
regarding $\b$ extraction from $\brpsik$ channel is different from
\cite{guetta}; the reason is explained later.

\vskip 5pt

In the SM, the matrix element of the effective Hamiltonian for
$\bbrpsik$ ($b \r c\bar{c} s$ at the quark level) is a combination of
tree and penguin contributions, given by
\bea
\label{hsm}
\bra J/\psi K_S|H_{\rm SM}|\overline{B_d}\ket 
& = & -\frac{G_F}{\sqrt 2} \left(A_{\rm SM}^{\rm tree} + 
A_{\rm SM}^{\rm peng}\right), ~~ {\rm where} \n,\\
A_{\rm SM}^{\rm tree} & = & V_{cb} V^*_{cs} (C_1 + \xi C_2) \apsi
~~~ (\xi = 1/N_c), \\
A_{\rm SM}^{\rm peng} & = & -V_{tb} V^*_{ts} (C_3 + \xi C_4
+ C_5 + \xi C_6 + C_7 + \xi C_8 + C_9 + \xi C_{10}) \apsi \n. 
\eea
Here $C_i$'s are the Wilson coefficients of the operators $O_1$-$O_{10}$
defined as in \cite{bu-fl} and evaluated at the factorization scale $\mu=
m_b$, and  
\bea
\label{apsi} 
\apsi & = & f_{J/\psi} F_0^{B\r K} f(m_B, m_K, m_{J/\psi}), \\
{\rm with}~~f(x,y,z) & = & \sqrt{x^4+y^4+z^4-2x^2y^2-2z^2y^2-2x^2z^2} \n.
\eea 
For numerical evaluation, we take the $J/\psi$ decay constant
$f_{J/\psi} = 0.38$ GeV \cite{ns} and the $B\rightarrow K$ decay form
factor $F_0^{B\r K} = 0.42$ \cite{bsw}.

\vskip 5pt

The $\lambda'$-induced interactions contribute to $b \r c\bar{c} s$
via slepton-mediated tree-level graphs. The matrix element of the
effective $\Rsl$ Hamiltonian is given by
\bea 
\label{hr}
\bra J/\psi K_S|H_\Rsl|\overline{B_d}\ket & = &
-\frac{1}{4} u^R_{222}~ \xi~ \apsi, \\
\label{u}
{\rm where,}~~ u^R_{jnk} &= & \sum_{i=1}^3
\frac{\lambda^{\prime *}_{in3} \lambda'_{ijk}}{2 \mslsq}. 
\eea 
Due to the QCD dressing to the above operator, the expression in
Eq.~(\ref{hr}) should be multiplied by a factor $\sim 2$ 
at the scale $m_b$ \cite{singer}, which we have taken into account in
all our numerical calculations.

\vskip 5pt

The presence of $\Rsl$ terms modifies the expression for CP
asymmetry in the $\brpsik$ channel  in the following way. 
The key parameter of course is 
\be 
\ljpsi = e^{-2i\b} {{\langle J/\psi K_S|\bar{B}_d\rangle} 
\over{{\langle J/\psi K_S|{B}_d\rangle}}}
 \equiv e^{-2i\b} {\bar{A} \over A}, 
\ee
\be
\label{a}
{\rm where,} ~~  A = A_{\rm SM} (1+ r e^{i\dphi} e^{i\dalfa})
~~[{\rm with}~ r = A_{\Rsl}/A_{\rm SM}],
\ee
\be
\label{abar}
\bar{A} = A_{\rm SM} (1+ r e^{-i\dphi} e^{i\dalfa}).  
\ee
In the above expressions, $A_{\rm SM}$ and $A_{\Rsl}$ denote the
signed magnitudes of the SM and $\Rsl$ amplitudes
respectively\footnote{$A_{\rm SM}$ contains both the tree and penguin
amplitudes of the SM. Notice that the CKM factors in Eq.~(\ref{hsm})
are both real to a very good approximation. We also neglect the strong
phase difference, expected to be small, that may appear between the
tree and the $I = 1$ part of the electroweak penguin amplitudes. Among
the different penguin contributions, only the dominant QCD part is
included in our numerical estimates.}, and we have ignored the overall
phases in the expressions of $A$ and $\bar{A}$. Here $\dphi$ and
$\dalfa$ are relative weak and strong phases between the SM and $\Rsl$
diagrams. The mixing induced CP asymmetry is given by
\be
\label{acpm} 
\acpmix \equiv {{2 {\rm Im} \ljpsi} \over {1+|\ljpsi|^2}} 
 = -{{2\rho}\over {1+\rho^2}} \sin({2\beta +\zeta}), 
\ee
\be 
\label{rho}
{\rm where,}~~ \rho = {|\bar{A}| \over |A|} = 
\left({{1+r^2+2r\cos(-\dphi + \dalfa)}\over
{1+r^2+2r\cos(\dphi + \dalfa)}}\right)^{1/2}, 
\ee
\be 
{\rm and,}~~ \zeta = 
\tan^{-1} \left(\frac{2r\sin \dphi (\cos \dalfa + r \cos
\dphi)}{1+r^2\cos(2\dphi)+2r\cos \dphi\cos\dalfa} \right).  
\ee
The direct CP asymmetry is given by  
\be
\label{acpd}
\acpdir \equiv \frac{1-|\ljpsi|^2}{1+|\ljpsi|^2}
= -\frac{2r \sin \dphi \sin \dalfa}{1+r^2+2r \cos\dphi \cos\dalfa}. 
\ee
The time-dependent CP asymmetry, 
\be
a_{\rm CP}(t) = {{B(\tpsik) - B(\tbarpsik)}\over 
{B(\tpsik) + B(\tbarpsik)}}, 
\ee   
is given by 
\be
\label{acpt} 
a_{\rm CP}(t) = \acpdir \cos(\Delta m_d t) + \acpmix \sin(\Delta m_d t). 
\ee
After time integration, one obtains
\be
\label{acptot} 
a_{\rm CP} = \frac{1}{1+x^2} \left[\acpdir + x \acpmix \right]; 
~~{\rm where}~ x = (\Delta M/\Gamma)_{B_d, B_s}.  
\ee

\vskip 5pt

Side by side, a measurement of direct CP asymmetry in the $\brpsikpl$
channel, which is the charged counterpart of $\brpsik$, yields
important information about new physics. The asymmetry is defined by 
\be 
\acppl = {{B(\brpsikpl) - B(\brpsikmi)}\over 
{B(\brpsikpl) + B(\brpsikmi)}}.  
\ee
To a good approximation, $\acppl = \acpdir$ (see Ref.~\cite{fm1} for
details). CLEO has measured \cite{cleo1} 
\be
\label{acppl} 
\acppl = (-1.8 \pm 4.3 \pm 0.4)\%. 
\ee

\vskip 5pt

The operators that mediate $\brpsik$ can have isospin $I$ either 0 or
1. In fact, one can write the effective Hamiltonian in the $I = 0$ and
$I = 1$ pieces in a model independent manner \cite{fm1}. In the SM the
$I = 1$ contribution suffers a dynamical suppression. Recall that a
sizable $\acpdir$ necessarily requires a large strong phase difference
(see Eq.~(\ref{acpd})), which can result only from the interference
between $I = 0$ and $I = 1$ amplitudes of comparable magnitude. In the
SM, the former is far more dominant than the latter. As a result,
$\acpdir$ in the SM is vanishingly small. In some extensions beyond
the SM, the $I = 1$ piece may be slightly enhanced. A large
enhancement however requires the presence of large rescattering
effects, which is not a likely scenario \cite{fm1}. In the present
context, new physics contributes only to the $I = 0$ sector by
inducing a set of slepton mediated tree diagrams in the $b \r
c\bar{c}s$ channel, for various combinations of $\lambda'$
couplings. Thus in the absence of any possible enhancement of the $I =
1$ part, the most likely scenario is that the strong phase difference
($\dalfa$) is still vanishingly small, and so is the direct CP
asymmetry. Yet, as we will see below, there is a significant impact of
non-zero $\dphi$ on the extraction and correct interpretation of $\b$.

\vskip 5pt

Following Eq.~(\ref{rho}), $\rho$ is identically equal to unity when
$\dalfa = 0$. Thus new physics can contaminate CP asymmetry only
through
\be 
\label{zetaspecial}
\zeta = 
\tan^{-1} \left(\frac{2r\sin \dphi (1 + r \cos
\dphi)}{1+r^2\cos(2\dphi)+2r\cos \dphi} \right).
\ee

\vskip 5pt

To determine the maximum allowed size of $r$, we have to find out the
experimental constraints on $\lambda^{\prime
*}_{i23}\lambda'_{i22}$. The best constraints come from the
measurements of $\brphikplmi$ branching ratio, which we will derive in
this paper. The essential formalism \cite{fm2} is described
below. First note that due to the isospin structure of the
interaction, the above product couplings contribute both to $b \r
c\bar{c} s$ (i.e. $\bbrpsik$ and $\brpsikplmi$) and $b \r s\bar{s} s$
(i.e. $\bbrphik$ and $\brphikplmi$) at tree level. While for the
former the diagram is slepton mediated, for the latter it is sneutrino
mediated. For simplicity we assume that both the slepton and sneutrino
are degenerate. Again note that for $\brphikplmi$ the leading SM
diagram is a penguin, while the $\Rsl$ interaction, as mentioned
before, proceeds at the tree level. The SM and $\Rsl$ effective
Hamiltonians for $\brphikmi$ lead to the following amplitudes:
\bea 
\label{hpsm}
\bra \phi K^-|H'_{\rm SM}|B^-\ket & = & 
\frac{G_F}{\sqrt 2} V_{tb} V^*_{ts}\times\nonumber\\ 
&{ }&\left[(C_3 + C_4)(1+\xi) + C_5 + \xi C_6 
-0.5\{C_7 + \xi C_8 + (C_9 + C_{10})(1+\xi)\}\right] 
\aphi, \\
\label{hpr}
\bra \phi K^-|H'_\Rsl|B^-\ket & = & 
-\frac{1}{4} \left(d^R_{222} + d^L_{222}\right) \xi~ \aphi,
\eea 
where $\aphi$ may be obtained analogously to the way $\apsi$ was
determined via Eq.~(\ref{apsi}), and 
\be 
d^R_{jkn} =  \sum_{i=1}^3
\frac{\lambda^{\prime *}_{in3} \lambda'_{ijk}}{2 \msnsq}, ~~~
d^L_{jkn} =  \sum_{i=1}^3
\frac{\lambda^{\prime *}_{inj} \lambda'_{i3k}}{2 \msnsq}.
\ee 
The most recent measurements of $\brphikplmi$ branching ratio are 
\be 
\label{brphik}
B (\brphikplmi) = 
\left\{
 \begin{array}{ll} 
\left(7.7 {}^{+1.6}_{-1.4} \pm 0.8\right) \times 10^{-6}
~~({\rm BaBar}~\cite{babar2}),\\
\left(13.9 {}^{+3.7}_{-3.3} {}^{+1.4}_{-2.4}\right) \times 10^{-6}
~~({\rm Belle}~\cite{belle2}),\\
\left(5.5 {}^{+2.1}_{-1.8} \pm  0.6\right) \times 10^{-6} 
~~({\rm CLEO}~\cite{cleo2}).
\end{array}
 \right.  \ee 
Now for simplicity we assume that only one of $d^R_{222}$ and
$d^L_{222}$ is non-zero, i.e. we do not admit unnatural cancellations
between them. Since we are interested to put bounds on
$\lambda^{\prime *}_{i23}\lambda'_{i22}$, we assume $d^R_{222}$ to be
non-zero.  We further simplify the situation by assuming that only one
combination, say the one corresponding to $i = 3$, is non-zero,
i.e. the exchanged scalar is a tau-sneutrino.  We also observe that
the weak phase associated with this product coupling is totally
arbitrary. In other words, $\lambda^{\prime *}_{323}\lambda'_{322}$
can take either sign, and $\dphi$ is completely unconstrained. Note
that the SM prediction for the $\brphikplmi$ varies in a wide range
$(0.7 - 16)\times 10^{-6}$ (see Table I of Ref.~\cite{cleo2}). To
appreciate how much new physics effect we can accommodate, we assume
that the SM contribution is close to the lower edge of the above
range, and then saturate the 2$\sigma$ CLEO upper limit in
Eq.~(\ref{brphik}) {\em entirely} by $\Rsl$ interactions. This 
way we obtain\footnote{While deriving the bound in
Eq.~(\ref{lpbound}), we have multiplied the effective Hamiltonian in
Eq.~(\ref{hpr}) by a QCD enhancement factor of 2, as was done with
Eq.~(\ref{hr}).}  
\be 
\label{lpbound}
|\lambda^{\prime *}_{323}\lambda'_{322}| 
~\ltap ~1.5 \times 10^{-3}, 
\ee
for an exchanged tau-sneutrino mass of 100 GeV. In fact, the above
constraint is valid for any lepton family index $i$. This is the best
constraint on the above combination\footnote{Notice that the
individual limits on $\lambda'_{i23}$ and $\lambda'_{i22}$ have been
extracted from squark mediated processes assuming a mass of 100 GeV
for whichever squark is involved \cite{review}. While a 100 GeV
slepton is very much consistent with all current data, the lower limit
on a generic squark mass is presently pushed up to around 300 GeV from
direct searches at Fermilab. Therefore, the $\lambda'$ limits derived
from squark mediated processes should be properly scaled while
comparing them with those extracted from slepton exchanged diagrams.},
which we have derived for the first time in this paper.

\vskip 5pt

To translate the limit in Eq.~(\ref{lpbound}) into a limit on $r$, we
need to decide in which regularization scheme $A_{\rm SM}$ will be
computed. If the Wilson coefficients are computed in the 't
Hooft-Veltman (HV) scheme\footnote{We have adapted the Wilson
coefficients from Table 26 of Ref.~\cite{bu-fl}. The uncertainties in
evaluating those coefficients arise from regularization scheme
dependence, choice of $\Lambda^5_{\bar{\rm MS}}$, the factorization
scale $\mu$, and the long distance corrections. The Wilson
coefficients we have used have been computed using 
$\Lambda^5_{\bar{\rm MS}} = 225$ MeV, $\mu = \bar{m}_b (m_b) = 4.4$
GeV and $m_t = 170$ GeV.}, $r$ lies in the range 
\be 
\label{r}
-0.3 ~\ltap~ r ~\ltap ~0.3.
\ee
(There is no qualitative change in the result if we use other schemes 
like Naive Dimensional Regularization (NDR).)
It is the negative value of $r$, which corresponds to a negative value
of the $\Rsl$ product coupling, that has got an interesting
implication in the extraction of $\b$. As an illustration, putting $r
= - 0.3$ and $\dphi = 90^\circ$ (say), it follows from
Eq.~(\ref{zetaspecial}) that $\zeta \sim -33^\circ $. 
Since our choices of $\lambda'$ couplings do not contribute to
$\Delta m_d$ or any other observables that go into the SM fit, the
latter still implies $\b \sim 22^\circ$ (see
Eq.~(\ref{smfit})), which we accept as the `true' value of $\b$. As a
result, the mixing induced CP
asymmetry in the $\brpsik$ channel, as conceived via Eq.~(\ref{acpm}),
now becomes $\sim 0.2$, as opposed to the SM expectation $\sim
0.7$. The crucial point is that a sizable negative $\zeta$ tends to
cancel the `true' $2\b$ in the argument of the sine function in the
expression of $\acpmix$. This example demonstrates that $\brpsik$ need
not be a `gold-plated' channel for the determination of $\b$. Rather,
it could provide a window for new physics to manifest. Note that no
drastic assumption, like large rescattering effects, etc, parametrized
by a large $\sin \dalfa$, was required to arrive at the above
conclusion. In Figure 1, we demonstrate the variation of $\sin 2\b$
with $r$ for different values of $\dphi$ and fixed $\dalfa =
0$. The $\dphi = 0$ curve corresponds to the SM reference value $\sin
2\b = 0.7$. The minimum $\sin 2\b$ we obtain in our scenario is 
\be 
\label{rp-bound}
\sin (2\b)_{\rm min} = 0.2, 
\ee
which should be compared with Eqs.~(\ref{l-bound}) and
(\ref{mssm-bound}). 

\begin{figure}[htb]
\centerline{
\psfig{file=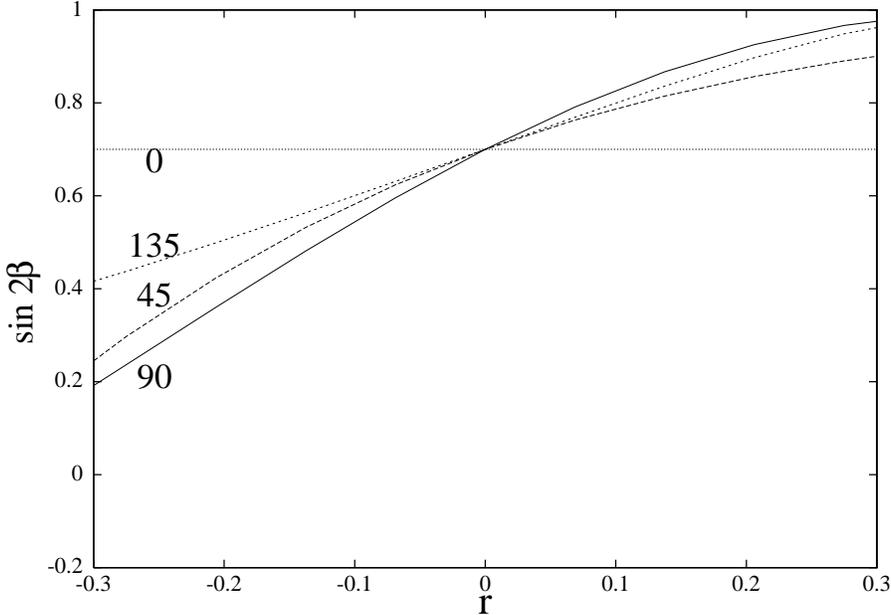,width=12cm,angle=270}
}
\caption{\small \sf Variation of $(\sin 2\beta)_{J/\psi K_S}$ as a
function of $r$ and the weak phase difference $\Delta \phi$ (defined
in the text). The values of $\Delta \phi$ (in degree) are indicated
adjacent to the lines. The strong phase difference $\Delta \alpha$ is
set to zero.}
\label{fig:figure1}
\end{figure}

\vskip 5pt

If we admit a fine-tuned situation $d^R_{222} = - d^L_{222}$ in
Eq.~(\ref{hpr}), then the $\brphikplmi$ branching ratio constraints
will not apply. But this product coupling will contribute to
$B_s$-$\bar{B}_s$ mixing on which there is only an experimental lower
limit. If we assume that in future the $B_s$-$\bar{B}_s$ mixing value
settles, say, close to its present lower limit, then one requires
$|\lambda^{\prime *}_{323}\lambda'_{322}| ~\sim ~2.7 \times 10^{-3}$
(via sneutrino mediated box graphs \cite{br}) to saturate the entire
mixing by $\Rsl$ interaction. This is a conservative approach, as for
a larger mixing one needs a larger value of the $\Rsl$ product
coupling. With the above value of the product coupling, it is possible
to arrange an even larger negative $\zeta$, than obtained using
Eq.~(\ref{r}), that can completely cancel the `true' $2\b$ inside the
sine function of $\acpmix$, which renders the latter almost
zeroish. It must be admitted, though, that such a delicate fine-tuning
is a very unlikely scenario.

\vskip 5pt

So far we have worked putting $\dalfa = 0$, i.e. by turning off any
possible strong phase difference between the interfering amplitudes
that might arise from long distance effects. Our approach has been a
conservative one. However, leaving aside any theoretical argument, one
can put an experimental constraint on $\dalfa$ for a given value of
$r$ and $\dphi$.  To do this we first take $r = -0.3$ and $\dphi =
90^\circ$ as a reference point, which yielded the minimum $\sin 2\b$
in our model.  We then put these values in Eq.~(\ref{acpd}) and
constrain $\dalfa$ from the 2$\sigma$ lower limit of the CLEO
measurement of direct CP asymmetry in Eq.~(\ref{acppl}). We obtain
$|\dalfa| ~\ltap ~11^\circ$. This should be seen as an experimental
constraint on long distance contributions for given values of the
other parameters. Indeed, the BaBar and Belle collaborations can
measure the sine and cosine time profiles of CP asymmetry in
Eq.~(\ref{acpt}). The hadronic machines, on the other hand, are
sensitive only to the time-integrated $\acp$ (see
Eq.~(\ref{acptot})). In the latter, the presence of a non-zero
$\acpdir$ can further disturb the extraction of $\b$ which resides
inside $\acpmix$.

\vskip 5pt

The experimental branching ratio $B(\brpsik) = (9.5 \pm 1.0) \times
10^{-4}$ \cite{gao} is a factor of three to four higher than what one
obtains from pure short distance effects involving naive
factorization \cite{ruckl}. We do not attempt to solve this puzzle in
this paper. We only make a remark that if the long distance
contributions for short distance SM and $\Rsl$ operators behave
similarly, then the long distance effects on $\acp$ will be
minimal. Since the SM and $\Rsl$ operator structures are different, a
separate study in this hitherto unexplored area is required. We do not
get into those details in this paper.

\vskip 5pt

A few comments on a previous analysis \cite{guetta} of $\Rsl$ effects
on $\brpsik$ are now in order. The bounds on $\Rsl$ couplings used in
\cite{guetta} are $|\lambda^{\prime *}_{i23}\lambda'_{i22}| ~\ltap
~1.4 \times 10^{-4}$ $(m_{\tilde{d}_R}/100 ~{\rm GeV})^2$ derived in
\cite{agashe}. As a result of such a strong constraint, the amplitude
of $\brpsik$ changes only marginally. The above analysis led to a
conclusion that the determination of $\b$ via $\brpsik$ was rather
robust. In this paper, we counter this conclusion on the following
ground. We notice that the bounds derived in \cite{agashe} are
`basis-dependent', whose meaning is explained below. If only one
$\lambda'$ coupling in the weak basis is non-zero, more than one such
coupling becomes non-zero in the mass basis, which are related to one
another by the CKM elements. This way one can generate
flavour-changing neutral currents (FCNCs) either in the down quark
sector or in the up quark sector, depending on whether the CKM mixing
is in the down sector or in the up sector respectively. Strong
constraints on the $\lambda'$ couplings emerge if one considers FCNCs
completely in the down sector. If one chooses the other extreme,
namely FCNCs entirely in the up sector, the constraints are not that
tight. Since the choice of this basis is completely arbitrary, we
prefer not to use such basis-dependent bounds in our analysis mainly
because the conclusion depends heavily on this choice.  Moreover, if
one duly scales the exchanged down-squark mass to 300 GeV in view of
its collider constraints, the limit used in \cite{guetta} gets relaxed
by an order of magnitude, and becomes closer to our limit in
Eq.~(\ref{lpbound}).  A general study of CP violating $B$ decays in
nonleptonic modes in supersymmetric models with broken $R$-parity can
also be found in Ref.~\cite{jang-lee}.

\vskip 5pt

The couplings $\lambda^{\prime *}_{i23}\lambda'_{i22}$ we have used in
our analysis to reproduce a low $(\b)_{J/\psi K_S}$ may have
nontrivial impact on other processes as well. If a correlation is
observed among the phenomena ocurring in these processes, it will
certainly provide a strong motivation for the kind of new physics
interactions we have advocated in our paper. These benchmark tests are
listed below:
\vskip 2pt

\noindent {\bf 1.}~ The SM amplitude of $B_s$-$\bar{B}_s$ mixing
involves $V_{tb} V^*_{ts}$, and is real to a very good
approximation. Hence we can expect the $b \r c\bar{c}s$ decays of the
$B_s$ meson (e.g. $B_s \r D_s^+ D_s^-$, $B_s \r J/\psi \phi$) to be CP
conserving. But as mentioned before, the $\lambda^{\prime
*}_{i23}\lambda'_{i22}$ couplings can contribute to $B_s$-$\bar{B}_s$
mixing through slepton mediated box graphs, which can interfere with
the SM diagram. The same combinations, we have seen before, also
affect the $b \r c\bar{c} s$ decay amplitude. The magnitude and the
weak phase of the above product coupling needed to produce a low CP
asymmetry in $\brpsik$ channel would necessarily give rise to a
sizable CP asymmetry in the $B_s \r D_s^+ D_s^-$ and $B_s \r J/\psi
\phi$ modes \cite{bdk}. Note that neither $R$-parity conserving SUSY
nor any other minimal flavour violating models can induce CP violation
in the latter channels. In some sense, therefore, its observation will
provide a necessary test for our scenario. The non-observation, on the
other hand, will rule out our explanation of low $\acp$ in
$\brpsik$. In any case, to observe CP violation in $B_s$ decays, we
have to wait till the second generation $B$ factories, namely the
LHC-b and BTeV, start taking data.

\vskip 2pt

\noindent {\bf 2.}~ We have put constraints on the magnitude of the
$\lambda^{\prime *}_{i23}\lambda'_{i22}$ couplings from the
experimental $\brphikplmi$ branching ratio. In fact, the CP asymmetry
in its neutral counterpart, namely the $\brphik$ channel, is again
expected to be proportional to $\sin 2\b$ in the SM. Now since the SM
amplitudes for $\brpsik$ and $\brphik$ are quite different, it is
quite apparent that in new physics inspired scenario, the CP
asymmetries measured in those two processes could yield values of $\b$
not only different from the `true' $\b$ but also different from each
other \cite{guetta}. In other words, $\sin (2\b)_{J/\psi K_S} \neq
\sin (2\b)_{\phi K_S} \neq \sin (2\b)_{\rm SM ~fit}$. Again, to verify
these non-equalities, we have to depend on the results from the second
generation $B$ factories.

\vskip 2pt

\noindent {\bf 3.}~ It has been pointed out in Ref.~\cite{cdk} that
the same $\lambda^{\prime *}_{i23}\lambda'_{i22}$ couplings can
successfully explain the $B \r \eta' K$ anomaly \cite{cleo-eta'}. It
is noteworthy that the magnitude and phase of the above couplings
required to produce a low $(\b)_{J/\psi K_S}$ explains the anomaly by
enhancing the SM branching ratio of $B \r \eta' K$ to its experimental
value. Note again that none of the minimal flavour violating models
can do this job.

\vskip 2pt

\noindent {\bf 4.}~ It has been claimed that $\sin 2\b$ can be
determined very cleanly from the branching ratio measurements of the
rare decays $K^+ \r \pi^+ \nu \bar{\nu}$ and $K_L^0 \r \pi^0 \nu
\bar{\nu}$ \cite{buchalla}. These processes, unlike those previously
mentioned, will not be affected by our choice of $\Rsl$
couplings. Comparison of $\beta$ extracted from these rare $K$ decay
processes with those obtained from several $B$ decay channels may
offer a powerful tool for probing physics beyond the SM.


\vskip 5pt 
 
To conclude, we have demonstrated that the distinctly lower central
value of the BaBar measurement of CP asymmetry in the $\brpsik$ channel
can be explained in models of supersymmetry with broken $R$-parity. In
the process, we have derived new upper limits on the relevant $\Rsl$
couplings from the experimental $\brphikplmi$ branching ratio. It should
be admitted though that the ability of the $\Rsl$ interactions to
lower $(\b)_{J/\psi K_S}$ is indeed shared by a few minimal flavour
violating models, e.g. MSSM or mSUGRA. Thus if the disagreement
between $(\b)_{J/\psi K_S}$ and $(\b)_{\rm SM ~fit}$ persists, this
will signal new physics no doubt, but just with this single piece of
information one cannot distinguish betwen the different models. What
makes our scenario special is that it can do certain other jobs what
the minimal flavour violating models cannot. We have outlined them
above. Some of these tests can be carried out only in the second
generation $B$ factories. These tests can either boost our scenario or
can rule it out.

\section*{Acknowledgments}
The work of AD has been supported by the DST, India (Project No.\
SP/S2/k01/97) and the BRNS, India (Project No.\ 37/4/97 - R \& D
II/474).  AK's work has been supported by the BRNS grant
2000/37/10/BRNS of DAE, Govt.\ of India, and by the grant F.10-14/2001
(SR-I) of UGC, India. We thank A. Raychaudhuri for reading the
manuscript.

\end{document}